# Modulation of ferromagnetism in (In,Fe)As quantum wells via electrically controlled deformation of the electron wavefunctions

Le Duc Anh[1], Pham Nam Hai[1,2], Yuichi Kasahara[3],
Yoshihiro Iwasa[3,4] and Masaaki Tanaka[1]

[1]*Department of Electrical Engineering and Information Systems,The University of Tokyo, 7-3-1 Hongo, Bunkyo-ku, Tokyo 113-8656, Japan*
[2]*Department of Physical Electronics, Tokyo Institute of Technology,2-12-1 Ookayama, Meguro, Tokyo 152-0033, Japan*
[3]*QPEC & Department of Applied Physics, The University of Tokyo, 7-3-1 Hongo, Bunkyo-ku, Tokyo 113-8656, Japan*
[4]*RIKEN Center for Emergent Matter Science, Wako 351-0198, Japan*

We demonstrate electrical control of ferromagnetism in field-effect transistors with a trilayer quantum well (QW) channel containing an ultrathin *n-type* ferromagnetic semiconductor (In,Fe)As layer. A gate voltage is applied to control the electron wavefunctions $\varphi_i$ in the QW, such that the overlap of $\varphi_i$ and the (In,Fe)As layer is modified. The Curie temperature is largely changed by 42%, whereas the change in sheet carrier concentration is 2 - 3 orders of magnitude smaller than that of previous gating experiments. This result provides a new approach for versatile, low power, and ultrafast manipulation of magnetization.

Recently, the change of magnetic properties using an external gate voltage in field effect transistor (FET) structure was demonstrated in carrier-induced ferromagnetic semiconductors (FMSs) [1,2,3,4,5,6,7,8] and ferromagnetic ultrathin metal films [9,10,11,12,13,14], which is expected to reduce the power consumption of spin devices. Commonly in these experiments, although the ferromagnetic channels are very thin (<5nm), the two-dimensional (2D) quantization is smeared out because of the carriers' low coherency. The Curie temperature ($T_C$) of these thin films monotonically depends on the carrier density $n$, which is varied by the effect of electrical gating, as predicted for three-dimensional (3D) ferromagnets [1,2,3]. A very large change in the sheet carrier density $n_{sheet}$ ($\Delta n_{sheet}$ = $10^{13}$-$10^{14}$ cm$^{-2}$) is typically required to generate a pronounced change in the magnetic properties, which in practice is difficult, and consumes a lot of energy ($E \propto \Delta n_{sheet}^2$). Furthermore, the response time is limited by the device's capacitance and the carriers' transit time from outside electrodes to the channel, which is at shortest picoseconds.

In 2D ferromagnetic systems such as FMS quantum wells (QWs), however, the physics is different, as illustrated in the top panel of Fig. 1(a): Due to the quantization of the density of states, $T_C$ does not change with the carrier density unless the Fermi level has reached the next subband [15]. Instead, theoretical studies suggested that $T_C$ is very sensitive to the shape of the carrier wavefunction $\varphi_i(z)$ of the occupied subband $i$, as described by the equation $T_C \propto \sum_i \int_{FMS} |\varphi_i(z)|^4 dz$ [16,17,18,19], where $z$ is the direction normal to the interface and the integral is performed inside the ferromagnetic layer in the QW (the fourth power is of great importance because it does *not* represent the carrier density, which is the square of the wavefunction, but represents the distribution of carriers in the ferromagnetic layer). Therefore, as illustrated in the bottom panel of Fig. 1(a), by using the gate voltage to deform the shape of the wavefunctions $\varphi_i$ of the QW, one can effectively control the magnetic properties *without the need of changing the carrier concentration*. This method has two important advantages over the conventional method: First, since no extra charge is needed, the charging energy of the FET's capacitor, which determines the power consumption for modulating magnetization, can be greatly reduced. Second, instead of travelling hundreds of nm-long paths as in the conventional method, electron carriers are redistributed over only a few nm in the QW. Therefore, the modulation speed can be as fast as ~ 100 femtoseconds [20,21]. Another feature of this method is the ability to tune the relation between $T_C$ and $V_G$ at will by specifying the appropriate position and thickness of the ferromagnetic layer inside the QW [16,17,18]. However, experimental demonstration of such functionality has never been reported due to the lack of a good ferromagnetic QW so far.

Here we demonstrate such a control of ferromagnetism of 2D ferromagnetic QW via deforming the carrier wavefunction, using electrical gating in a FET structure. The system used in this study is a semiconductor InAs-based QW that contains a thin n-type FMS (In,Fe)As layer. Among the many types of FMSs, (In,Fe)As is unique and promising because it is the only reliable n-type FMS among III-V based semiconductors [22]. In this material, Fe atoms in the isovalent state $Fe^{3+}$ replace In atoms and thus play only the role of local magnetic moments. Electron carriers, which are independently supplied by defects or donors, interact strongly with these local Fe magnetic moments through the *s-d* exchange interaction to induce ferromagnetism with a $T_C$ of up to several tens of Kelvin with a modest electron density (~ $6 \times 10^{18}$ cm$^{-3}$), and there is plenty of room for improvement [19,22,23,24,25,26]. We have observed that the electron carriers in (In,Fe)As reside in the conduction band and exhibit high coherency [19,24]. In fact, magnetic circular dichroism (MCD) spectroscopy has confirmed the QSE in QWs consisting of InAs/(In,Fe)As/InAs trilayers with a total thickness as large as 40 nm, in which the electron wavefunctions smoothly extend throughout the trilayers [19,25]. These features of the n-type FMS (In,Fe)As make it possible to demonstrate the electrical control of ferromagnetism via wavefunction manipulation.

We prepared a QW structure (sample A) consisting of InAs (2 nm)/(In$_{0.94}$,Fe$_{0.06}$)As (8 nm)/InAs (5 nm) on an AlSb buffer, all grown on a semi-insulating GaAs (001) substrate by molecular beam epitaxy (left panel in Fig. 1(b)) [27]. Because of the large conduction-band offset at the bottom InAs/AlSb interface (1.35 eV) [28], the electrons are confined to the top InAs/(In,Fe)As/InAs trilayer and form quantized electronic states. The QSE is evident in the blue shift of the MCD spectrum of sample A [27]. We used an electric-double-layer-type field effect transistor (FET) structure to control the carrier wavefunctions in the magnetic trilayer QWs, as illustrated in Fig. 1(c). The sample was etched into a $50 \times 200$ μm$^2$ Hall bar using standard photolithography and ion milling. A side-gate electrode (G), a reference electrode (R) and several electrodes (source S, drain D, and electrodes numbered 1-3) for transport measurements were formed via the electron-beam evaporation and lift-off of an Au (50 nm)/Cr (5 nm) film. The side-gate pad (G), the reference pad (R) and the (In,Fe)As channel were covered with electrolyte (DEME-TFSI) to form the FET structure. Other regions of the device were separated from the electrolyte by an insulating resist. As illustrated in the inset of Fig. 1(c), when a positive $V_G$ is applied, ions in the electrolyte accumulate at the surface of the semiconductor channel and form an electric double-layer capacitor, which changes the potential and electron density in the InAs/(In,Fe)As/InAs trilayer QW.

The transport and magnetic properties of the trilayer QW were characterized by Hall measurements [27]. The Hall resistance of (In,Fe)As always consists of a large n-type normal Hall resistance (NHR) and a much smaller anomalous Hall resistance (AHR) that is proportional to the magnetization. The small AHR (~3% of the total Hall resistance) can be obtained by subtracting the negative slope of the NHR at a high magnetic field (1 T) from the raw Hall-resistance data. The large contribution of the NHR in the Hall resistance data allows us to accurately estimate the electron density from the Hall measurements. $T_C$ was estimated by monitoring the temperature dependence of the remanent AHR. Figure 1(d) presents the anomalous Hall resistance (AHR) of the trilayer QW in device A at $V_G$ = 0 V at various temperatures, in which clear ferromagnetic hysteresis is observed at low temperatures. The inset presents the remanent AHR vs. temperature $T$, indicating a $T_C$ of 24 K.

Figure 2(a) shows the temperature dependence of the resistance ($R$ - $T$) measured between terminals 1 and 3 of device A ($R_{13}$) using a 4-terminal method. $R_{13}$ systematically decreases (increases) when a positive (negative) $V_G$ is applied, as expected. Figure 2(b) shows $n_{sheet}$ in the InAs/(In,Fe)As/InAs trilayer QW at various $V_G$ values estimated by Hall measurements at 4.2 K. The $n_{sheet}$ of device A (red circles) changes from $6.06 \times 10^{12}$ cm$^{-2}$ (at $V_G = 0$ V) to $8 \times 10^{12}$ cm$^{-2}$ (at $V_G = 6$ V) or to $5.52 \times 10^{12}$ cm$^{-2}$ (at $V_G = -3$ V). The $\Delta n_{sheet}$ is small, probably because of the existence of a large number of surface states at the top InAs layer, which tends to pin the Fermi level.

Figures 3(a) and (c) show the evolution of the AHR of device A at 15 K when $V_G$ is adjusted as follows: 0 V → 0.5 V → 6 V → 0 V and 0 V → –3 V → 0 V, respectively. At $V_G = 6$ V and –3 V, the hysteresis in the AHR - $H$ characteristics disappears, whereas it recovers after $V_G$ is returned to 0. These results demonstrate that the ferromagnetism of the (In,Fe)As thin film can be reversibly controlled by electrical gating. Figures 3(b) and (d) show the remanent AHR $vs.$ $T$ of device A at various positive and negative $V_G$ values, respectively, and illustrate the evolution of $T_C$. As summarized in Fig. 4(a), the $T_C$ of the (In,Fe)As layer of device A (black circles) decreases from its initial value of 24 K ($V_G = 0$ V) to 17 K ($V_G = 6$ V) or 14 K ($V_G = -3$ V).

Compared with previous electrical-gating experiments [1-14], the present results for device A exhibit two distinct features. First, very effective control of $T_C$ is realized (the largest change in $T_C$, $[T_C(V_G)-T_C(0)]/T_C(0)$, is -42% at $V_G = -3$ V) despite a very small change in $n_{sheet}$ (the corresponding $\Delta n_{sheet} = -5.4 \times 10^{11}$ cm$^{-2}$). This $\Delta n_{sheet}$ is 2 - 3 orders of magnitude smaller than the change of the $n_{sheet}$ that was required in almost all the previous gating experiments (the only exceptions whose $\Delta n_{sheet}$ are comparable with the present study are CdMnTe QW[6] and GeMn quantum dots[7], in which, however, the ferromagnetism was turned off by depleting carriers in the ferromagnetic channels). This means that the energy for charging the ferromagnetic channel is dramatically reduced by a factor of $10^{-4}$-$10^{-6}$ while maintaining the same effect of modulating the ferromagnetism. Second, $T_C$ varies non-monotonically with $n_{sheet}$, reaching a maximum near $V_G = 0.5$ V and decreasing at both positive and negative $V_G$. This is characteristic of 2D FMS QWs, in which the overlap of the electron wavefunctions and the (In,Fe)As layer determines the single $T_C$ value of the entire system. Fig. 4(b) presents the changes in the potential profile (blue curves) and the electron wavefunctions (yellow shapes) of the QW in device A at $V_G = 0$, 6, and $-3$ V, which were obtained through self-consistent calculations [27]. The calculated electron density distribution in the QW versus $V_G$ is summarized in Fig. 4(c). The integral of the electron density distribution over the entire QW yields $n_{sheet}$. The overlap of the wavefunctions and the local Fe magnetic moments in the (In,Fe)As layer is largest at $V_G = 0.5$ V and decreases as the wavefunctions move towards the ends of the trilayer QW at both positive and negative $V_G$, leading to the observed behavior of $T_C$. These results clearly indicate the feasibility of controlling the ferromagnetism via wavefunction manipulation and confirm the intrinsic electron-induced ferromagnetism of n-type FMS (In,Fe)As.

In a 2D FMS structure, $T_C$ is given by [16,17,19]

$$T_C^{2D} = \frac{S(S+1)}{12} \frac{A_F^{2D} J_{sd}^2}{k_B} \frac{m^*}{\pi\hbar^2} N_{Fe} \sum_{E_i < E_F\,(In,Fe)As} \int |\varphi_i(z)|^4 dz \quad (1)$$

Here, $z$ is the growth direction, $S$ is the spin angular momentum of an Fe atom (=5/2), $m^*$ is the effective electron mass at the Γ point, $k_B$ is the Boltzmann constant, $J_{sd}$ is the s-d exchange interaction constant, $A_F^{2D}$ is the Stoner enhancement factor in 2D structures [29], and $N_{Fe}$ is the Fe atom density. The calculated $T_C^{2D}$ values (red squares in Fig. 4(a)) exhibit good agreement with the experimental results. It is worth noting that a semi-classical approach based on a modified 3D mean-field Zener model, in which the quantum size effect (QSE) is ignored, cannot explain the behavior of $T_C$ observed in the present experiment [27]. From the calculations of $T_C^{2D}$, the effective s-d exchange interaction energy $N_0\alpha$ of the (In,Fe)As QW was estimated to be 3.6 eV. This effective $N_0\alpha$ value was calculated from the value of $(A_F^{2D})^{1/2} J_{sd}$, which implicitly includes the enhancement effect associated with electron-electron interactions in 2D structures.

Finally, we demonstrate the ability to control the relation between $T_C$ and $V_G$ by modifying the QW structure. Another sample and its corresponding FET device were prepared (device B) using the same procedure but a modified QW structure: InAs (1.6 nm)/(In$_{0.94}$,Fe$_{0.06}$)As (4 nm)/InAs (2.4 nm)/Si-doped InAs (4 nm) (right panel of Fig. 1(b)). Si donors ($5 \times 10^{18}$ cm$^{-3}$) in the bottom 4-nm InAs layer supply electrons to the QW and attract the 2D electron wavefunctions towards the bottom AlSb side. This feature, together with the reduced thickness (4 nm) of the (In,Fe)As layer, creates a situation in which the wavefunctions and Fe magnetic moments are separated when $V_G = 0$; however, their overlap increases as a positive $V_G$ is applied. As shown in Fig. 2(b), device B exhibits a higher $n_{sheet}$ ($1.08 \times 10^{13}$ cm$^{-2}$ at $V_G = 0$) than device A but a weak dependence of $n_{sheet}$ on $V_G$, similar to device A. Figure 3(e) shows the evolution of the AHR-$H$ characteristics of device B at 25 K for $V_G = 0$ V → 2 V → 0 V, and Fig. 3(f) summarizes the remanent AHR *vs*. *T* curves at various $V_G$ values. The sign of AHR in device B is positive, which is opposite to that of device A. The QWs in device A and B are different in the inversion asymmetry (due to the space charge potential of activated Si donors in device B) and $n_{sheet}$, both of which could be the origin of the different AHR sign [30,31]. Positive AHR has also been observed in bulk-like (In,Fe)As samples doped with Be that have high electron density[22]. Thus, at this stage we suggest that the sign change of AHR is probably due to the difference in $n_{sheet}$, although a more elaborate study is needed to clarify the mechanism. As can be seen in Fig. 3(e), unlike device A, the ferromagnetism of the (In,Fe)As layer in device B is enhanced by applying positive $V_G$, as expected, and $T_C$ increases from 27 K (at $V_G$ = 0 V) to a maximum of 35 K (at $V_G$ = 2 V) or to 30 K (at $V_G$ = 4 V), as illustrated in Fig. 4(a) (blue circles). The $T_C$ values calculated by the 2D mean-field Zener model for device B are plotted as green diamonds, which also show a good agreement with the experimental values (see SM for details of the calculations). Thus, using the appropriate QW structure, one can either increase or decrease the $T_C$ of the (In,Fe)As layer in QWs with the same device configuration. This high degree of freedom in controlling the ferromagnetism is an important advantage, which is attractive for device applications.

In summary, we have demonstrated the electrical control of ferromagnetism in (In,Fe)As QWs by manipulating the overlap between the 2D wavefunctions and the (In,Fe)As layer. The $T_C$ was largely changed by 10 K ($\Delta T_C / T_C$ = 42%), whereas the change in $n_{sheet}$ is 2 - 3 orders of magnitude smaller than that of previous gating experiments. The behavior of $T_C$ was quantitatively explained by the Zener mean-field model for FMS QWs. We have also demonstrated the ability to customize the $T_C$ - $V_G$ relation by modifying the QW structure. These results confirm the intrinsic electron-induced ferromagnetism of n-type FMS (In,Fe)As and open new possibilities of using this material in spintronic devices, as well as provide a new approach for versatile, low power, and ultrafast manipulation of magnetization.

This work was partially supported by Grants-in-Aids for Scientific Research (Nos. 23000010, 25000003, 2474022, and 257388), including the Specially Promoted Research, the Project for Developing Innovation Systems of MEXT, and the Strategic International Collaboration Research Program (SICORP-LEMSUPER) from JST. L. D. A. acknowledges the support from the JSPS Fellowship for Young Scientists and the Program for Leading Graduate Schools (MERIT). P. N. H acknowledges supports from the Yazaki Memorial Foundation for Science and Technology, the Murata Science Foundation, and the Toray Science Foundation.

**Figure captions**

FIGURE 1. (a) Schematic illustration of the two methods: Controlling $T_C$ by carrier density as in previous works (top panel) and by carrier wavefunction as in the present study (bottom panel). In ferromagnetic QW, $T_C$ does not change proportionally with the carrier density as in 3D cases, which is due to the step-like change of the density of states. (b) Sample structures of device A (left panel) and device B (right panel). In both samples, the QW is the top InAs/(In$_{0.94}$,Fe$_{0.06}$)As/InAs trilayer structure. The Fe concentration is fixed at 6%. In device B, the bottom 5-nm InAs layer in the QW is doped with Si ($5\times 10^{18}$ cm$^{-3}$). (c) Schematic structure of the FET device with electrolyte (DEME-TFSI) between the gate (G), the reference electrode (R) and the trilayer QW. The source (S) and drain (D) electrodes and the electrodes numbered 1, 2, and 3 are used for transport

measurements. (d) AHR of device A at various temperatures. The inset shows the temperature dependence of the remanent AHR, which indicates a $T_C$ of 24 K.

FIGURE 2. (a) Temperature dependence of the $R_{13}$ of device A at various $V_G$ values. (b) $n_{sheet}$ values of device A (red circles) and device B (blue circles) at 4.2 K as functions of $V_G$.

FIGURE 3. (a), (c) Evolution of the AHR of device A measured at 15 K when adjusting $V_G$ as follows: 0 V → 0.5 V → 6 V → 0 V and 0 V → -3 V → 0 V, respectively. The insets present the AHR data near the origin, which exhibit clear changes in hysteresis characteristics. (b), (d) Temperature dependence of the remanent AHR of device A at various $V_G$ values. The colored arrows indicate the $T_C$ values of device A at each $V_G$. (e) Evolution of the AHR of device B measured at 25 K when adjusting $V_G$: 0 V → 2 V → 0 V. The inset presents the AHR data near the origin. (f) Temperature dependence of the remanent AHR of device B at various $V_G$ values. The colored arrows indicate the $T_C$ values of device B at each $V_G$.

FIGURE 4. (a) Experimental $T_C$ values of device A (black circles) and device B (blue circles), respectively, as functions of $V_G$. Red squares and green diamonds represent the $T_C^{2D}$ values calculated using the 2D mean-field Zener model for device A and B, respectively, which exhibit good agreement with the experimental results. (b) Calculated potential profiles (blue curves), electron wavefunctions (yellow shapes), and electron-

density distributions (red curves, corresponding to the right-hand vertical axes) of the QW of device A at $V_G$ = 0, 6 and -3 V, from top to bottom. The Fermi levels (green dashed lines) and quantized energy levels (black lines, in the eV unit) in the trilayer QWs are shown. (c) Calculated evolution of the electron density distribution in the trilayer QW of device A at 4.2 K when $V_G$ is varied from -3 V to 6 V. The pink regions correspond to the (In,Fe)As layer.

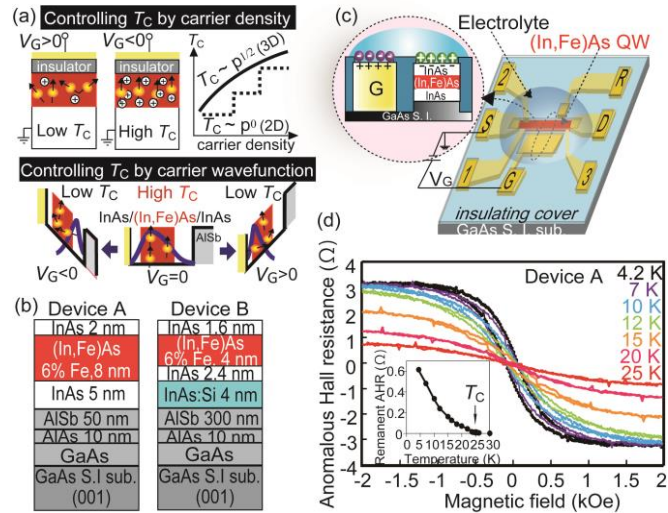

**FIGURE 1.** Anh *et al.*

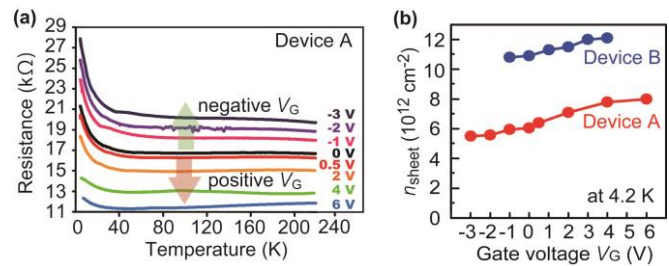

**FIGURE 2.** Anh *et al.*

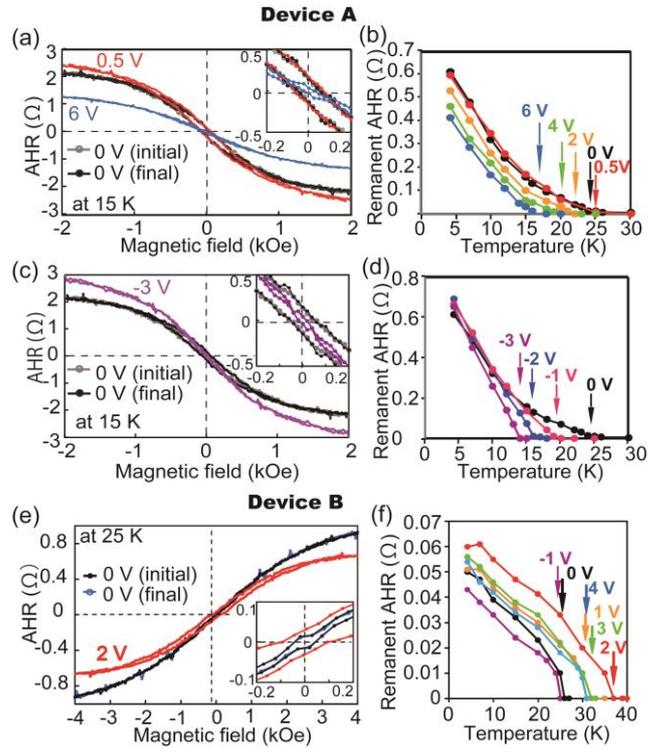

**FIGURE 3.** Anh *et al.*

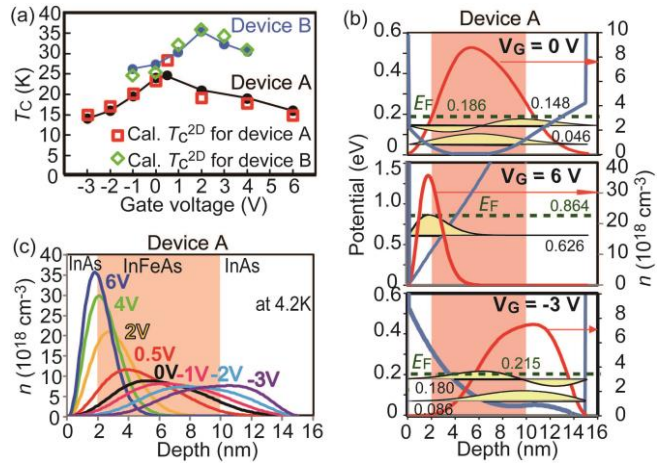

**FIGURE 4.** Anh *et al.*

# Supplemental Material

# Modulation of ferromagnetism in (In,Fe)As quantum wells via electrically controlled deformation of the electron wavefunctions

Le Duc Anh[1], Pham Nam Hai[1,2], Yuichi Kasahara[3],
Yoshihiro Iwasa[3,4] and Masaaki Tanaka[1]

[1]Department of Electrical Engineering and Information Systems, The
University of Tokyo, 7-3-1 Hongo, Bunkyo-ku, Tokyo 113-8656, Japan
[2]Department of Physical Electronics, Tokyo Institute of Technology, 2-12-1
Ookayama, Meguro, Tokyo 152-0033, Japan
[3]QPEC & Department of Applied Physics, The University of Tokyo, 7-3-1
Hongo, Bunkyo-ku, Tokyo 113-8656, Japan
[4]RIKEN Center for Emergent Matter Science, Wako 351-0198, Japan

1. **Sample preparation**

The samples were epitaxially grown by molecular beam epitaxy (MBE). The growth temperature ($T_S$) was 550℃ for the GaAs and AlAs layers, and the first 25 nm of the AlSb layer; 470℃ for the remaining AlSb layer; and 236℃ for the (In,Fe)As layer and the top InAs layer. For the bottom InAs layer of the trilayer QW, $T_S$ was 400℃ for device A and 250℃ for device B. The Fe concentration in the (In,Fe)As layer was 6%. The bottom 5-nm InAs layer in the QW of device B was doped with Si at a concentration of $5 \times 10^{18}$ cm$^{-3}$. *In situ* reflection high-energy electron diffraction (RHEED) revealed bright and streaky patterns during the growth of the AlSb barrier and the trilayer InAs/(In,Fe)As/InAs QW, indicating a good two-dimensional growth mode.

We calibrated the thickness of the InAs/(In,Fe)As/InAs QWs by cross-sectional transmission electron microscopy (TEM) characterizations. TEM images of the devices A and B are shown in Fig. S1. Different contrast in the two TEM images is due to the different thickness (along the incident electron beam direction) of the two specimens for the TEM observation. In device B's TEM specimens, which is thin enough, a lattice image is observed and good zinc-blende structure is confirmed in all the epitaxial layers. The thicknesses of the QWs are 14.8 nm and 12.1 nm in device A and B, respectively. The (In,Fe)As layer cannot be distinguished from the top and bottom InAs layers.

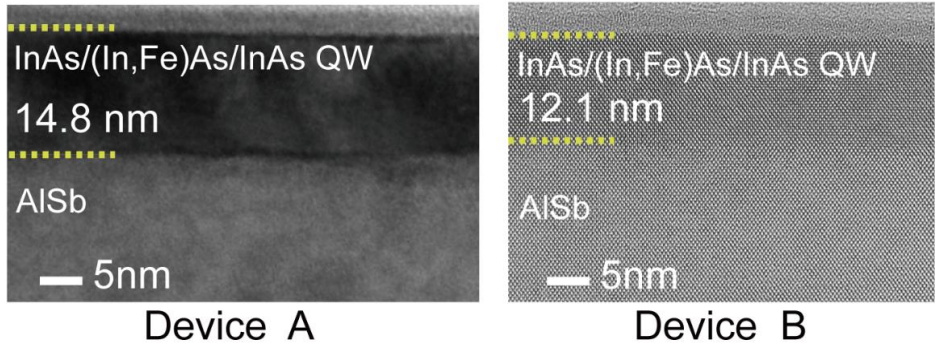

**FIG S1. Cross-sectional TEM image of the (In,Fe)As tri-layer QWs in device A and B.**

## 2. Characterization of magnetic properties
### 2.1 Hall measurements

Our FET devices were mounted on a cold finger of a He-flow cryostat. Transport and Hall-effect measurements were performed on our field effect transistor (FET) devices. We used the odd function of the Hall resistance *vs.* magnetic field ($H$) data to remove any contribution from the magnetoresistance, which is an even function of the magnetic field. Figure S2(a) and (b) shows the Hall resistances (HR) and anomalous Hall resistance (AHR) of device A and B, respectively, at temperatures below and above the Curie temperatures ($T_C$). The AHR was obtained by subtracting the linear normal Hall resistance (NHR) whose slope was determined at high magnetic field (1 T at 4.2 K and 0.2 T at 30 K). The anomalous Hall coefficient is negative in device A but positive in device B.

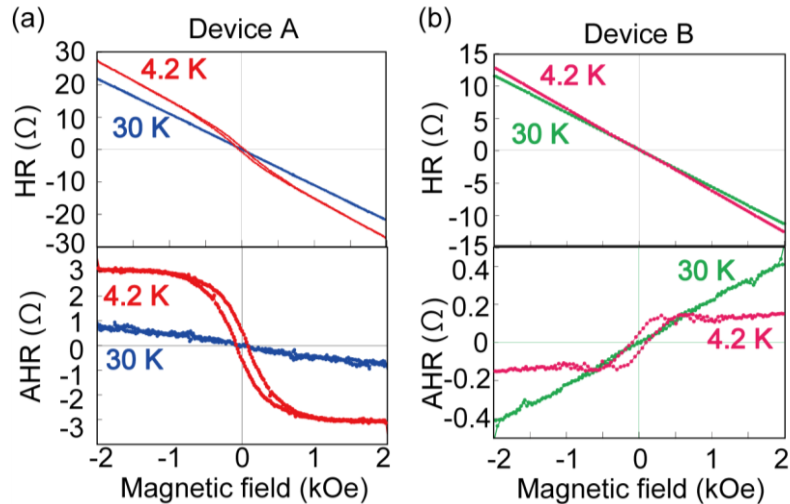

**FIG. S2. Hall resistances (HR, upper panel) and anomalous Hall resistance components (AHR, lower panel) of (a) device A and (b) device B at 4.2K and 30K, respectively, at $V_G = 0$V. The anomalous Hall resistance was obtained by subtracting the linear normal Hall resistance whose slope was determined at high magnetic field (1 T at 4.2 K and 0.2 T at 30 K).**

$T_C$ can be determined by monitoring the remanent AHR or by Arrott plots of the AHR – $H$ curves ($H$ is the magnetic field applied normal to the film plane) with increasing temperature. Figure S3(a), (b), (c) show the Arrott plots (AHR$^2$ – $H$/AHR plots) of device A at $V_G$ = 0 V, 6 V and -3 V, respectively. The $T_C$ estimated by the Arrott plots are the same as those obtained by monitoring the remanent AHR at various temperatures. This confirms the validity of the latter method, which is used in the main text.

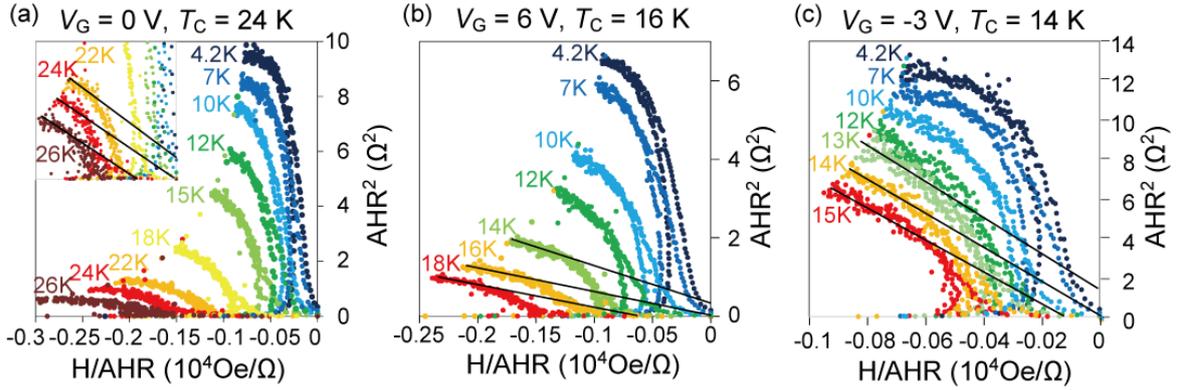

**FIG. S3.** Arrott plots of (In,Fe)As QW in device A at (a) $V_G$ = 0 V, (b) $V_G$ = 6 V, (c) $V_G$ = -3 V. The inset in (a) shows the magnified plot near $T_C$.

## 2.2 Magnetic circular dichroism (MCD) spectroscopy

In reflection MCD, we measure the difference in optical reflectivity between right ($R_{\sigma+}$) and left ($R_{\sigma-}$) circular polarizations, which is induced by the spin splitting of the band structure at a magnetic field $B$ of 1 T applied perpendicular to the film plane. The MCD intensity is expressed as follows: $\text{MCD} = \frac{90}{\pi} \frac{(R_{\sigma+} - R_{\sigma-})}{2R} \propto \frac{90}{\pi} \frac{1}{2R} \frac{dR}{dE} \Delta E$, where $R$ is the reflectivity, $E$ is the photon energy, and $\Delta E$ is the spin-splitting energy (Zeeman energy) of the material. Because of the $dR/dE$ term, a MCD spectrum shows peaks corresponding to the optical transitions at critical point energies of the band structure. At the same time, the MCD intensity is proportional to the magnetization $M$ of the measured material, because $M \propto \Delta E$ in FMSs. Therefore, MCD measurements give information of both the magnetization and the electronic structure of the material.

Figure S4(a) and (b) shows the MCD spectra measured around the energy of the critical point $E_1$ of (In,Fe)As (2.61 eV) for sample A and sample B, respectively, at 5 K and a perpendicular magnetic field of 1 T. The MCD spectra of both samples show structures that can be deconvoluted into two Lorentzian curves (blue dotted curves). These two Lorentzian curves in each device correspond to the optical transitions between the valence band and the 1st and 2nd quantized levels in the InAs/(In,Fe)As/InAs trilayer quantum wells (QWs). The energy difference between the two quantized levels is 120 meV and 145 meV for device A and B, respectively.

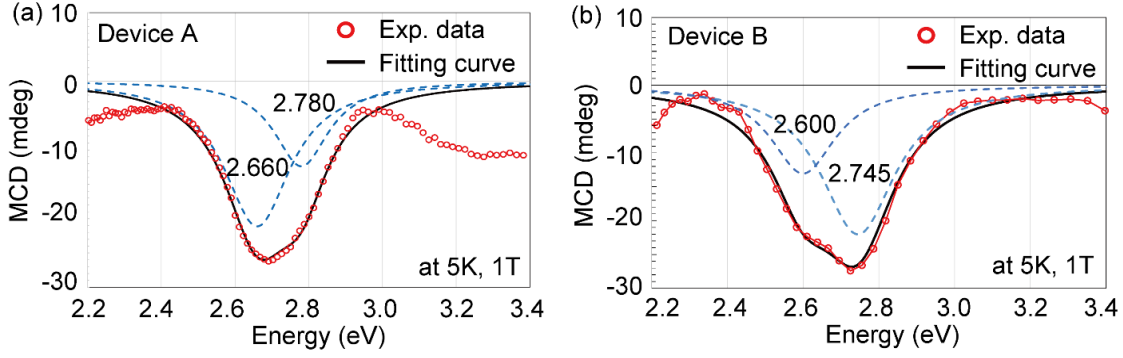

**FIG. S4. (a) and (b) MCD spectra of sample A and sample B (red circles) performed in a reflection setup at 5 K with a magnetic field of 1 T applied perpendicular to the film plane. The fitting curves (black curves) are summed up from the two Lorentzian curves (blue dotted curves) that correspond to the optical transitions between the valence band and the 1st and 2nd quantized levels of the conduction band. The numbers above the Lorentzian curves are the peak photon-energy values in eV.**

### 3. Calculations of quantum well potentials, wavefunctions and Curie temperatures
### 3.1 Calculation method

Self-consistent calculations using the Schrödinger equation (S1) and the Poisson equation (S2) were performed to obtain the potential profiles and carrier wavefunctions of the QWs.

$$\left(\frac{-\hbar^2}{2m^*}\frac{\partial^2}{\partial z^2} + V_{charge}(z) + V_{offset}(z) + V_{xc}(z) + V_{gate}(z)\right)\varphi(z) = E\varphi(z) \tag{S1}$$

$$\frac{\partial^2}{\partial z^2} V_{charge} = -e\frac{N_d(z) - \rho_e(z)}{\varepsilon} \tag{S2}$$

Here, $z$ is the growth direction, $V_{charge}$ is the space charge potential induced by ionized donors $N_d$ in the QW and electron carriers, $V_{offset}$ is the conduction band offset between InAs and AlSb at the $\Gamma$ point (1.35 eV), $V_{xc}$ is the exchange-correlation potential of electrons[S1], $V_{gate}$ is the potential induced by the gate voltage, and $\varphi(z)$ is the electron wavefunction. For the dielectric constant of (In,Fe)As, the value of InAs $\varepsilon = 12.37\varepsilon_0$ was used. $m^*$ is the effective mass of electrons at the $\Gamma$ point. In our recent paper [S2], we have estimated $m^*$ of 100-nm-thick (In,Fe)As thin films, which strongly depends on the electron density $n$ when $n$ is higher than $6\times10^{18}$ cm$^{-3}$, as shown in Fig. S5 (colored data points). In device A, $n_{sheet} = 6\times10^{12}$ cm$^{-2}$ is equivalent to a 3D electron density $n_{3D}$ of $1.5\times10^{19}$ cm$^{-3}$, while in device B, due to the Si doping, we have higher $n_{sheet} = 1\times10^{13}$ cm$^{-2}$ which is equivalent to $n_{3D} = 3.2\times10^{19}$ cm$^{-3}$. From the dependence of $m^*$ on $n$ in Fig. S5, we set $m^*$ at 0.08 $m_0$ in device A and 0.25 $m_0$ in device B (corresponding to the red and blue circle in Fig S5, respectively) in our calculations.

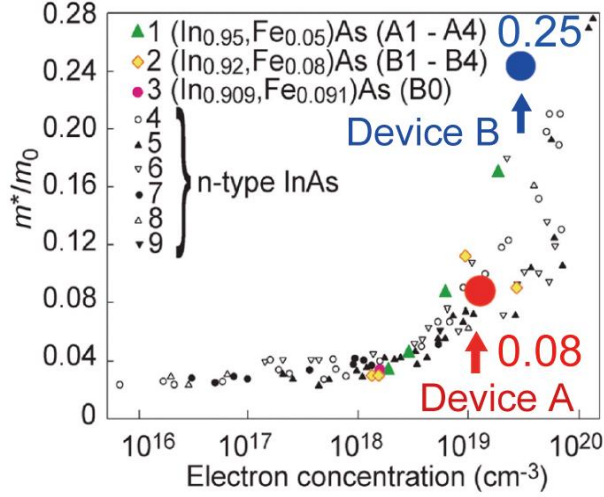

**FIG. S5. Dependence of the electron effective mass $m^*$ of (In,Fe)As on the electron density (small colored data points), comparing with that of n-type InAs (black and white data points) [from ref. S2]. The big red and blue circles with pointed arrows are the expected values for device A (0.08 $m_0$) and device B (0.25 $m_0$).**

The local electron density along the growth direction $\rho_e(z)$ was calculated as below, with the Fermi level $E_F$, quantized level energies $E_i$ and the Fermi-Dirac distribution function at low temperature simplified as the Heaviside function:

$$\rho_e(z) = \sum_{E_i<E_F} \frac{m^*}{\pi\hbar^2}(E_F - E_i)|\varphi_i(z)|^2 \tag{S3}$$

At $V_G = 0$ V, the Fermi level at the top InAs surface was assumed to be pinned at 50 meV [S3] above the bottom of the conduction band because of the surface states. At $V_G \neq 0$ V, the voltage drop $V_{gate}$ between the two ends of the QW was assumed to be $sV_G$ ($0<s<1$). The Fermi level at $V_G \neq 0$ V was determined by fitting the calculated $n_{sheet}$ to the $n_{sheet}$ values measured from the Hall effect. Therefore, $s$ was the only tunable parameter in our calculations and was obtained by matching the calculated $T_C$ to the experimental $T_C$.

Based on the mean-field Zener model for 2D FMS QW [S4,S5], $T_C$ is given by:

$$T_C^{2D} = \frac{S(S+1)}{12} \frac{A_F^{2D} J_{sd}^2}{k_B} \frac{m^*}{\pi \hbar^2} N_{Fe} \int_{FMS} |\varphi(z)|^4 dz \tag{S4}$$

Here, $S$ is the spin angular momentum of an Fe atom (=5/2), $k_B$ is the Boltzmann constant, $J_{sd}$ is the *s-d* exchange interaction constant, $A_F^{2D}$ is the Stoner enhancement factor in 2D structures, and $N_{Fe}$ is the Fe atom density. The original 2D mean-field Zener model was simplified by the assumption that the electron carriers occupy only the first quantized level. In order to describe for our (In,Fe)As QWs, we extended the model to the case of multiply occupied quantized levels. Thus $T_C$ was calculated as in equation (S5), which neglects the inter-level interaction effects:

$$T_C^{2D} = \frac{S(S+1)}{12} \frac{A_F^{2D} J_{sd}^2}{k_B} \frac{m^*}{\pi \hbar^2} N_{Fe} \sum_{E_i < E_F} \int_{(In,Fe)As} |\varphi_i(z)|^4 dz \tag{S5}$$

Note that the spin polarization in the QW was initially neglected in the calculations of the QW electronic structures but was perturbatively taken into account in the calculation of $T_C$.

In device A, ionized donors are the defects due to the low growth temperature in the top 10nm of the QW. The concentration $N_d = 1.2 \times 10^{19}$ cm$^{-3}$ was determined by fitting the calculated sheet carrier density $n_{sheet}$ of the QW to the $n_{sheet}$ value measured from the Hall effect. The *s* values ($s_+$, $s_-$) of device A obtained for positive $V_G$ and negative $V_G$ were 0.6 and 0.23, respectively; the asymmetry between these values originated from the asymmetry of the gating effect in our FET devices as can be seen in Fig. 2 in the main text.

In device B's case, ionized donors include donor-defects (concentration $N_d$) in the top 10 nm and Si (concentration $N_{Si}$) in the bottom 5 nm of the QW. In Fig. S6 we show the best-fit result of device B with $N_d = 1.8 \times 10^{19}$ cm$^{-3}$, $N_{Si} = 5 \times 10^{18}$ cm$^{-3}$, $m^* = 0.25\, m_0$, and the ($s^+$, $s^-$) = (0.81, 0.05).

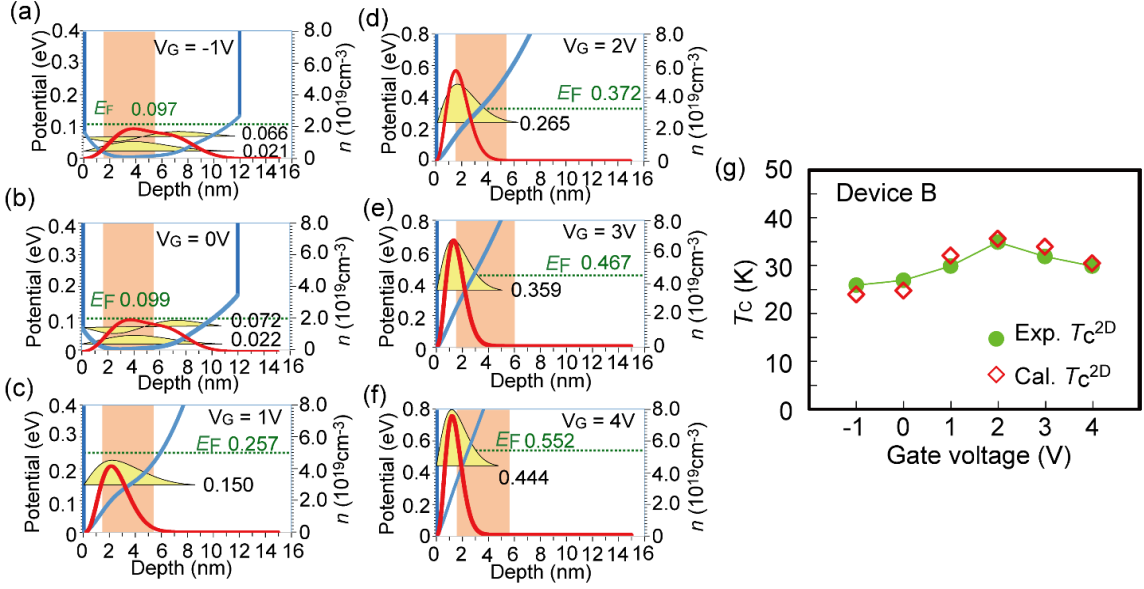

**FIG S6.** (a)-(f) Calculated QW potential (blue curves), wavefunctions (yellow curves) and electron density distribution (red curves) in device B for $V_G$ = -1, 0, 1, 2, 3, 4 V, respectively, with $N_d = 1.8 \times 10^{19}$ cm$^{-3}$, $N_{Si} = 5 \times 10^{18}$ cm$^{-3}$ and the ($s^+$, $s^-$) = (0.81, 0.05). (g) Experimental $T_C$ and those obtained by 2D mean-field Zener model (also shown in Fig. 4(a) in the main text).

### 3.2. Calculation of the Curie temperature ($T_C$) of the (In,Fe)As thin film in device A by the modified three-dimensional (3D) mean-field Zener model

In FMS thin films such as p-type Mn-based FMS (Ga,Mn)As, due to the strong disorder scatterings, the mean free path of carriers (holes) is shorter than the channel thickness and the two-dimensional quantization is smeared out even in the films as thin as 5 nm. Therefore, the electrical control of ferromagnetism by wavefunction manipulation cannot be realized for these FMS thin films. In such cases, a semi-classical approach by the modified three-dimensional (3D) mean-field Zener model was used, in which the quantum size effect (QSE) is ignored [S6].

Here, for comparison, we also calculated the potential profile and carrier distribution of in our trilayer QWs using the modified 3D mean-field Zener model. For the 3D case, we use only the Poisson equation. Similar to the calculations for the two-dimensional (2D) case described in the main text, the potential profiles were calculated from the potentials formed by space charges [electron carriers in the QW, donor antisite defects in the QW, and native acceptor defects ($N_A = 5 \times 10^{17}$ cm$^{-3}$) in the AlSb buffer], the conduction band offset, and the electron exchange-correlation potential at each gate bias voltage $V_G$. At $V_G = 0$ V, the Fermi level at the top InAs surface is assumed to be pinned at 50 meV above the conduction band bottom due to the surface states. At $V_G \neq 0$ V, the voltage drop between the two ends of the QW was assumed to be $sV_G$ ($0<s<1$). The Fermi level at $V_G \neq 0$ V was determined by fitting the calculated $n_{sheet}$ to the $n_{sheet}$ values measured by the Hall effect. For the 3D case, the donor antisite defects density $N_D$ obtained by fitting the calculated $n_{sheet}$ of the QW to the experimental value is $3.5 \times 10^{19}$ cm$^{-3}$.

The Curie temperature $T_C^{3D}$ was calculated by a modified 3D model used for the ultrathin GaMnAs channel [S6]:

$$T_C^{3D} = \int dz T_C[n(z), N_{Fe}] \int dz n(z)^2 / n_{sheet}^2 \quad (S6)$$

Here $T_C[n(z), N_{Fe}]$ is the local $T_C$ calculated by the 3D mean-field Zener model, and $n(z)$ is the electron density distribution along the growth direction $z$ inside the QW. Figure S7 shows the $T_C^{3D}$ values as a function of gate voltage ($V_G$), calculated for device A with $(s_+, s_-) = (0.6, 0.23)$ (red triangles) and $(1,1)$ (blue squares), respectively, where $s_+$ and $s_-$ are the $s$ parameters when $V_G$ is positive and negative, respectively. One can see that $T_C^{3D}$ cannot explain the experimental $T_C$ results (black circles), especially the decrease of the experimental $T_C$ at positive $V_G$, even when assuming that the whole gate voltage is applied only to the QW [i.e. the case of $(s_+, s_-) = (1,1)$]. In contrast, $T_C^{2D}$ calculated by the 2D modified Zener model can explain the experimental $T_C$ results, as shown in Fig. 4 in the main text.

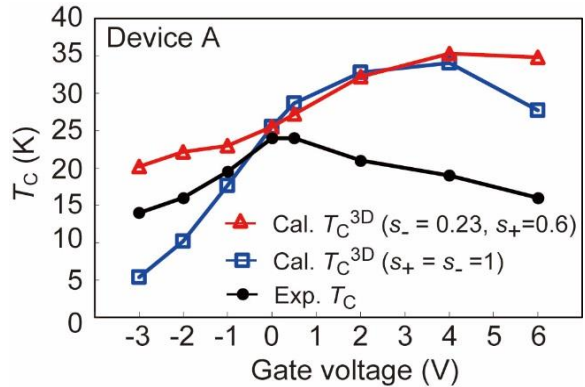

**FIG. S7. Curie temperature ($T_C$) of device A as a function of gate voltage ($V_G$). $T_C^{3D}$ calculated by the 3D mean-field Zener model with $(s_+, s_-) = (0.6, 0.23)$ (red triangles) and $(1,1)$ (blue squares) compared with the experimental $T_C$ values (black circles). $T_C^{3D}$ cannot explain the experimental results.**